\documentclass[english,aps,prb,twocolumn,superscriptaddress,citeautoscript]{revtex4-2}
\usepackage[latin9]{inputenc}
\usepackage{refstyle}
\usepackage{amsmath}
\usepackage{amssymb}
\usepackage{graphicx}
\usepackage{wasysym}

\makeatletter

\AtBeginDocument{\providecommand\subsecref[1]{\ref{subsec:#1}}}
\RS@ifundefined{subsecref}
  {\newref{subsec}{name = \RSsectxt}}
  {}
\RS@ifundefined{thmref}
  {\def\RSthmtxt{theorem~}\newref{thm}{name = \RSthmtxt}}
  {}
\RS@ifundefined{lemref}
  {\def\RSlemtxt{lemma~}\newref{lem}{name = \RSlemtxt}}
  {}

\usepackage{hyperref}

\makeatother

\usepackage{babel}
\begin{document}
\title{Dynamical Mean Field Theory of the Bilayer Hubbard Model with Inchworm
Monte Carlo}
\author{Dolev Goldberger}

\affiliation{School of Physics, Tel Aviv University, Tel Aviv 6997801, Israel}
\author{Yehonatan Fridman}
\affiliation{Department of Physics, NRCN, P.O. Box 9001, Beer Sheva 84190, Israel}
\affiliation{Department of Physics, Ben-Gurion University of the Negev, Beer-Sheva
84105, Israel}
\author{Emanuel Gull}
\affiliation{Department of Physics, University of Michigan, Ann Arbor, Michigan
48109, USA}
\author{Eitan Eidelstein}
\affiliation{Department of Physics, NRCN, P.O. Box 9001, Beer Sheva 84190, Israel}
\affiliation{School of Chemistry, Tel Aviv University, Tel Aviv 6997801, Israel}
\author{Guy Cohen}
\email{gcohen@tau.ac.il}

\affiliation{School of Chemistry, Tel Aviv University, Tel Aviv 6997801, Israel}
\affiliation{The Raymond and Beverley Sackler Center for Computational Molecular
and Materials Science, Tel Aviv University, Tel Aviv 6997801, Israel}
\begin{abstract}
Dynamical mean-field theory allows access to the physics of strongly
correlated materials with nontrivial orbital structure, but relies
on the ability to solve auxiliary multi-orbital impurity problems.
The most successful approaches to date for solving these impurity
problems are the various continuous time quantum Monte Carlo algorithms.
Here, we consider perhaps the simplest realization of multi-orbital
physics: the bilayer Hubbard model on an infinite-coordination Bethe
lattice. Despite its simplicity, the majority of this model's phase
diagram cannot be predicted by using traditional Monte Carlo methods.
We show that these limitations can be largely circumvented by recently
introduced Inchworm Monte Carlo techniques. We then explore the model's
phase diagram at a variety of interaction strengths, temperatures
and filling ratios.
\end{abstract}
\maketitle

\section{Introduction}

In strongly correlated materials quantum many-body physics shapes
the electronic properties. Such materials exhibit a wide variety of
unique and interesting phases and transitions \citep{morosan_strongly_2012}.
Many of these behaviors are thought to be beyond the predictive power
of standard simulation schemes like the density functional theory,
where electronic correlation is essentially treated as a static mean
field. Embedding methods like the dynamical mean field theory (DMFT)
\citep{georges_dynamical_1996}---which contains strong, but local,
correlations---have therefore emerged. For model systems, DMFT becomes
exact in certain infinite coordination number limits \citep{georges_hubbard_1992}.
It can also be used to introduce approximate strong correlation physics
into ab initio calculations made by, e.g., density functional theory
\citep{anisimov_first-principles_1997}.

DMFT can be phrased as a mapping between an extended model with many-body
interactions between orbitals in each unit cell; and a quantum impurity
problem representing the correlated orbitals in one unit cell, coupled
to an effective noninteracting bath that represents the rest of the
system. The coupling density between the impurity and bath, and the
connection between the impurity self-energy and the lattice self-energy,
are determined by a self-consistency condition \citep{georges_dynamical_1996}.
This means that to solve systems with multiple orbitals in each unit
cell, a method for solving multiorbital quantum impurity problems
is required. In this context, ``solving'' refers to the calculation
of Green's functions.

While several alternatives exist, the methods most often employed
by DMFT practitioners for general multiorbital problems are known
collectively as continuous-time quantum Monte Carlo (CT-QMC) algorithms
\citep{gull_continuous-time_2011}. These methods take perturbative
diagrammatic expansions and sum them to very high orders by Monte
Carlo integration over diagrams, with most calculations performed
in imaginary time on the Matsubara contour. Several different methods
exist, based on different expansions and with different regimes of
applicability; when these methods do break down, the breakdown typically
takes the form of a ``sign problem'' \citep{troyer_computational_2005},
such that the calculation becomes exponentially harder with decreasing
control parameter. For example, the interaction expansion (CT-INT)
\citep{rubtsov_continuous-time_2004} and auxiliary field (CT-AUX)
\citep{gull_continuous-time_2008} methods develop sign problems away
from half-filling. The hybridization expansion (CT-HYB) \citep{werner_continuous-time_2006,werner_hybridization_2006}
can work better in the presence of large, complicated interactions
\citep{gull_performance_2007}, but often develops sign problems when
the hybridization in the impurity model has large off-diagonal components.
More recently, we proposed an inchworm Monte Carlo method\citep{cohen_taming_2015}
based on the hybridization expansion, which can circumvent some of
the sign problems afflicting multiorbital DMFT calculations \citep{eidelstein_multiorbital_2020}.

One of the earliest models used to investigate multiorbital physics
within the framework of DMFT is the bilayer Hubbard model on the infinite
dimensional Bethe lattice \citep{moeller_rkky_1999}, which we present
in detail in Sec.~\ref{subsec:Model} below. While originally introduced
to study the relationship between short-ranged spin correlations and
the Mott metal--insulator transition, the model has reemerged over
the years: for example, more realistic variations on the model have
been used to simulate the metal insulator transition in vanadium dioxide
\citep{biermann_dynamical_2005,najera_resolving_2017,najera_multiple_2018}.
A two-dimensional version on a square lattice has been considered
as a toy model for unconventional superconductivity \citep{maier_pair_2011}
and realized experimentally using ultracold atoms in an optical lattice
\citep{gall_competing_2021}. Finally, extensions to hexagonal lattices
are currently of great interest in the study of layered 2D materials
\citep{xu_tunable_2022}.

The infinite-dimensional bilayer Hubbard model on the Bethe lattice
is exactly solvable within DMFT, given an exact solution for the corresponding
two-orbital impurity problem. Its phase diagram at half-filling, which
includes a Mott metal--insulator transition and an antiferromagnetic
regime, was investigated first by Hirsch--Fye Monte Carlo\citep{fuhrmann_mott_2006}
and shortly after by CT-INT \citep{hafermann_metal-insulator_2009}.
Notably, the 2D model on a square lattice, where DMFT is only approximate,
was also probed using an extension of DMFT with a discrete approximation
for the bath \citep{kancharla_band_2007}, and by a variational Monte
Carlo technique \citep{ruger_phase_2014}. The paramagnetic regime
of the doped system was recently investigated using CT-HYB \citep{fratino_doping-driven_2022},
by taking advantage of a specialized symmetry property to remove the
diagonal hybridizations; this suggested the existence of an interesting
pseudogap phase at intermediate doping.

Here, we show that the inchworm Monte Carlo method can be used as
an impurity solver for the bilayer Hubbard model within DMFT. The
robustness of the inchworm scheme to sign problems allows us to extend
the phase diagram beyond half-filling without any particular symmetry
restrictions, enabling exploration of ferromagnetic and antiferromagnetic
response. We use this to investigate the effect of doping on the metallic
and magnetic properties of the system, and analyze the results to
reveal simple short-ranged mechanisms behind much of the phenomenology.
The rest of this work is structured as follows: Sec.~\ref{sec:Theory}
briefly describes the model (\ref{subsec:Model}) and methodology
(\ref{subsec:Methodology}). In Sec.~\ref{sec:Results} we first
compare the CT-HYB impurity solver with its inchworm counterpart (\ref{subsec:solver-comparison}).
We then show how we determine phase transitions from our numerical
data (\ref{subsec:transitions}), and finally present and discuss
the phase diagram (\ref{subsec:phase-diagram}). In Sec.~\ref{sec:Conclusions}
we conclude and discuss future prospects.

\section{Theory\label{sec:Theory}}

\subsection{Model\label{subsec:Model}}

\begin{figure}
\includegraphics{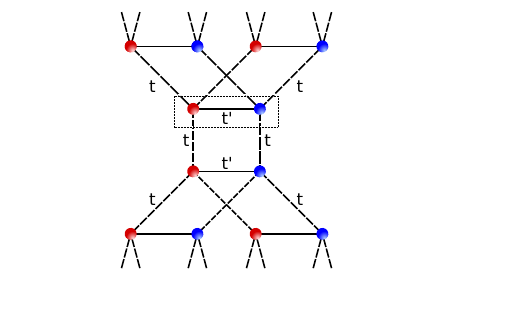}

\caption{Illustration: a small region in the bilayer Hubbard model on a Bethe
lattice with $Z=3$ (we consider the limit $Z\rightarrow\infty$).
Blue and red spheres, respectively, represent orbitals in the first
and second layer. Orbitals on adjacent sites within a layer are coupled
by hopping amplitudes $t$. Within each dimer or unit cell, one of
which is marked by a rectangular box, particles can hop between layers
with hopping amplitude.\label{fig:blh_bethe}}
\end{figure}
We define the bilayer Hubbard model on the Bethe lattice in the infinite
coordination number limit $Z\rightarrow\infty$. In Fig.~\ref{fig:blh_bethe}
we provide an illustration with $Z=3$. The model comprises two identical
Hubbard models, or layers, on Bethe lattices with hopping $t$, chemical
potential $\mu$ and interaction strength $U$. Each site in the first
layer is coupled to its counterpart in the second with hopping $t^{\prime}$.
The Hamiltonian is given by:
\begin{alignat}{1}
H & =-t\sum_{\left\langle ij\right\rangle \sigma}\left(a_{i\sigma}^{\dagger}a_{j\sigma}+b_{i\sigma}^{\dagger}b_{j\sigma}\right)+t'\sum_{i\sigma}\left(a_{i\sigma}^{\dagger}b_{i\sigma}+b_{i\sigma}^{\dagger}a_{i\sigma}\right)\nonumber \\
 & -\sum_{i\sigma}\mu\left(n_{i\sigma}^{\left[a\right]}+n_{i\sigma}^{\left[b\right]}\right)+U\sum_{i}\left(n_{i\uparrow}^{a}n_{i\downarrow}^{a}+n_{i\uparrow}^{b}n_{i\downarrow}^{b}\right).\label{eq:hamiltonian}
\end{alignat}
The subscript $\sigma\in\left\{ \uparrow,\downarrow\right\} $ denotes
spin, and $a^{\dagger}$($b^{\dagger}$) creates an electron in layer
$a$($b$). In the second line, $n_{i\sigma}^{a}\equiv a_{i\sigma}^{\dagger}a_{i\sigma}$
and $n_{i\sigma}^{b}\equiv b_{i\sigma}^{\dagger}b_{i\sigma}$. We
also define the ratio between the inter-layer and intra-layer hopping,
$\alpha\equiv t'/t$. To obtain a nontrivial infinite coordination
number limit, we rescale $t$ to $\frac{t}{\sqrt{Z-1}}$, then set
this to $1$ to define our unit of energy. This scaling is not applied
to $t^{\prime}$.

\subsection{Methodology\label{subsec:Methodology}}

The DMFT method provides a self consistent mapping from Eq.~(\ref{eq:hamiltonian})
onto a quantum impurity model \citep{georges_dynamical_1996}. In
the present model and for antiferromagnetic order, the self consistency
requirement takes on a particularly simple form \citep{moeller_rkky_1999,hafermann_metal-insulator_2009}:
\begin{equation}
\begin{aligned}\mathcal{\boldsymbol{G}}_{\sigma}^{-1}\left(i\omega_{n}\right) & \equiv\int_{0}^{\beta}\mathrm{d}\tau e^{i\omega_{n}\tau}\mathcal{\boldsymbol{G}}_{\sigma}^{-1}\left(\tau\right)\\
 & =\left(\begin{array}{cc}
i\omega_{n}+\mu & -\alpha t\\
-\alpha t & i\omega_{n}+\mu
\end{array}\right)-\boldsymbol{\Delta}_{\sigma}\left(i\omega_{n}\right),
\end{aligned}
\label{eq:self-consistency}
\end{equation}
where $\mathcal{\boldsymbol{G}}_{\sigma}\left(i\omega_{n}\right)$
is the Weiss field at Matsubara frequency $\omega_{n}$. Here the
hybridization function,
\begin{equation}
\boldsymbol{\Delta}_{\sigma}=t^{2}\boldsymbol{G}_{-\sigma}\left(i\omega_{n}\right),\label{eq:hyb_func}
\end{equation}
is obtained from the local Green's function at the (arbitrarily chosen)
site $i=0$, such that $a_{\sigma}=a_{0\sigma}$ and $b_{\sigma}=b_{0\sigma}$:
\begin{equation}
\boldsymbol{G}_{\sigma}\left(\tau\right)\equiv-\left(\begin{array}{cc}
\left\langle \mathrm{T}_{\tau}a_{\sigma}\left(\tau\right)a_{\sigma}^{\dagger}\left(0\right)\right\rangle  & \left\langle \mathrm{T}_{\tau}a_{\sigma}\left(\tau\right)b_{\sigma}^{\dagger}\left(0\right)\right\rangle \\
\left\langle \mathrm{T}_{\tau}b_{\sigma}\left(\tau\right)a_{\sigma}^{\dagger}\left(0\right)\right\rangle  & \left\langle \mathrm{T}_{\tau}b_{\sigma}\left(\tau\right)b_{\sigma}^{\dagger}\left(0\right)\right\rangle 
\end{array}\right).\label{eq:greens_function}
\end{equation}
Note that this assumes the spins in adjacent sites can be flipped
by antiferromagnetism, forming a bipartite lattice. A small symmetry
breaking term was also applied to the Hamiltonian in the initial DMFT
iteration when seeking antiferromagnetic solutions.

The impurity model to be solved is then defined by the effective action
\begin{equation}
\begin{aligned}S_{\mathrm{eff}} & =-\sum_{\sigma}\int_{0}^{\beta}\mathrm{d}\tau\mathrm{d}\tau^{\prime}\mathrm{d}\tau^{\prime}\mathbf{c}^{\dagger}\left(\tau\right)\cdot\boldsymbol{\mathcal{G}}_{\sigma}^{-1}\left(\tau-\tau^{\prime}\right)\cdot\mathbf{c}\left(\tau^{\prime}\right)\\
 & +U\int_{0}^{\beta}\mathrm{d}\tau\left(n_{\uparrow}^{a}\left(\tau\right)n_{\downarrow}^{a}\left(\tau\right)+n_{\uparrow}^{b}\left(\tau\right)n_{\downarrow}^{b}\left(\tau\right)\right),
\end{aligned}
\label{eq:impurity_problem}
\end{equation}
where $\mathbf{c}^{\dagger}\left(\tau\right)\equiv\left(a^{\dagger}\left(\tau\right),b^{\dagger}\left(\tau\right)\right)$,
$n_{\sigma}^{a}\left(\tau\right)\equiv a_{\sigma}^{\dagger}\left(\tau\right)a_{\sigma}\left(\tau\right)$
and $n_{\sigma}^{b}\left(\tau\right)\equiv b_{\sigma}^{\dagger}\left(\tau\right)b_{\sigma}\left(\tau\right)$.
This is a two-orbital impurity problem. For this particular problem,
at $\mu=-U/2$ the system becomes particle--hole symmetric, and CT-INT
works without a sign problem \citep{hafermann_metal-insulator_2009}.
The off-diagonal elements in Eq.~(\ref{eq:hyb_func}) lead to sign
problems in CT-HYB, which can be eliminated in the paramagnetic case
by transforming to the bonding/antibonding orbital basis, where $\boldsymbol{\Delta}_{\sigma}$
is diagonal. Neither algorithm works robustly in the general case.
Here, we employed an inchworm Monte Carlo algorithm based on CT-HYB,
but able to deal with the multiorbital sign problem emerging from
nondiagonal hybridization functions \citep{eidelstein_multiorbital_2020}.
At their core, inchworm algorithms are a resummation technique for
diagrammatic Monte Carlo methods. They were originally developed almost
a decade ago to address the dynamical sign problem for populations
dynamics in the real time hybridization expansion\citep{cohen_taming_2015}
and related expansions for the spin--boson model \citep{chen_inchworm_2017,chen_inchworm_2017-1,cai_numerical_2020,cai_fast_2022,yang_inclusion-exclusion_2021,cai_inchworm_2023},
but were soon extended to Green's functions\citep{antipov_currents_2017}
and used to perform real-time DMFT calculations \citep{dong_quantum_2017}.
In addition to a variety of applications in nonequilibrium physics
\citep{ridley_numerically_2018,ridley_lead_2019,ridley_numerically_2019,krivenko_dynamics_2019,kleinhenz_dynamic_2020,erpenbeck_revealing_2021,kleinhenz_kondo_2022,pollock_reduced_2022,erpenbeck_quantum_2023},
several extensions and generalizations have been aimed at making these
methods more useful as imaginary time impurity solvers \citep{li_interaction-expansion_2022,kim_pseudoparticle_2022,strand_inchworm_2023}.
Our present implementation includes several optimizations not present
in the previous work \citep{eidelstein_multiorbital_2020}, including
fast diagram summations\citep{boag_inclusion-exclusion_2018} and
a sparse representation of local propagators that takes advantage
of conserved quantum numbers. We plan to discuss these technical aspects
in detail in future work along with a public release of our code.

We note in passing that recent advances in tensor train methods as
a substitute for Monte Carlo techniques can also circumvent sign problems,
and may offer an alternative route to solving the impurity problems
appearing here \citep{nunez_fernandez_learning_2022,erpenbeck_tensor_2023}.
Novel decomposition schemes for finite-order diagrammatic methods
are another interesting development in this regard \citep{kaye_decomposing_2023}.

\section{Results\label{sec:Results}}

\subsection{Sign problem in CT-HYB and inchworm algorithm\label{subsec:solver-comparison}}

\begin{figure}
\includegraphics{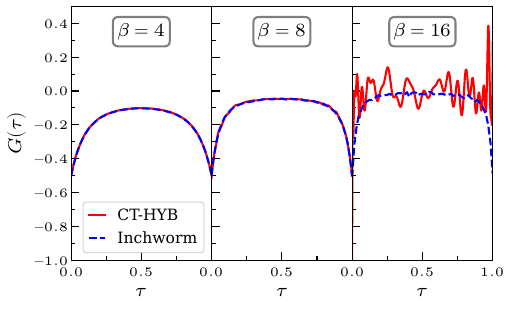}

\caption{Development of the sign problem in CT-HYB (solid red curves) as temperature
is decreased, and comparison to inchworm Monte Carlo (dashed blue
curves) at interaction strength $U=4$, $\alpha=1$ and half-filling.
The same computation time is used for each of the six simulations
shown.\label{fig:sign_problem_in_gf}}
\end{figure}

The DMFT algorithm consists of a sequence of iterations, in each of
which the impurity solver is used to compute the Green's function
of Eq.~(\ref{eq:impurity_problem}) for a hybridization function
obtained from the previous iteration or---in the first iteration---from
the noninteracting problem. In the CT-HYB impurity solver \citep{werner_continuous-time_2006},
generally a very efficient choice at anything but the smallest interaction
strengths \citep{gull_performance_2007}, sign problems appear when
the hybridization function develops large nondiagonal components.
To demonstrate this, consider Fig.~\ref{fig:sign_problem_in_gf},
where the Green's function is calculated for interaction strength
$U=4$ and tunneling amplitude ratio $\alpha=1$, at half-filling.
All simulations shown in the figure and in the rest of this subsection,
with both methods, were obtained with the same total amount of computer
time ($\sim150$ minutes on 56 CPU cores). We used the implementation
of CT-HYB \citep{shinaoka_continuous-time_2017} based on the ALPSCore
libraries \citep{gaenko_updated_2017}. While CT-HYB (solid red curves)
performs well at the higher temperatures $\beta=4$ and $\beta=8$,
it breaks down at $\beta=16$. The corresponding data from an inchworm
calculation \citep{eidelstein_multiorbital_2020} based on the same
expansion (dashed blue curves) remains reliable at the entire temperature
range shown.

\begin{figure}
\includegraphics{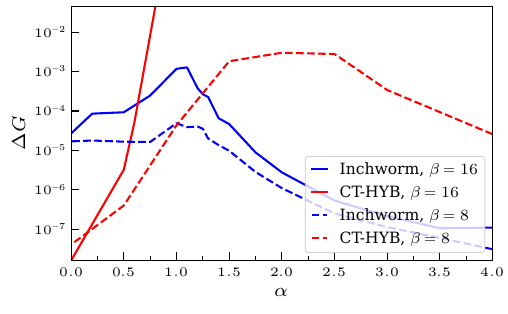}

\caption{Stochastic error estimates from inchworm (blue curves) and CT-HYB
(red curves) as a function of $\alpha$ at two temperatures (solid
and dashed curves). The interaction strength is set to $U=3$ and
the system is half-filled.\label{fig:sign_problem_error}}
\end{figure}

While the example in Fig.~\ref{fig:sign_problem_in_gf} is illustrative,
it is difficult to ascertain how typical it is without further exploring
the parameter space. To facilitate this it is convenient to define
a measure of the error, and we choose the jackknife variance over
$n=5$ independent realizations of the calculation with different
random seeds, averaged over imaginary time:
\begin{equation}
\Delta G\equiv\frac{1}{\beta}\int_{0}^{\beta}\mathrm{d}\tau\frac{1}{n\left(n-1\right)}\sum_{i=1}^{n}\left(G_{i}\left(\tau\right)-\bar{G}\left(\tau\right)\right)^{2}.
\end{equation}
In Fig.~\ref{fig:sign_problem_error} we plot this measure on a logarithmic
scale as a function of $\alpha$, for $U=3$ and two temperatures,
still at half-filling. Results from CT-HYB and inchworm are shown
in red and blue, respectively; while results at $\beta=16$ and $8$
are shown in solid and dashed curves, respectively. Once again, every
data point uses the same overall computer time as in Fig.~(\ref{fig:sign_problem_in_gf}).
CT-HYB is substantially more accurate at small $\alpha$, i.e. when
there is little or no dimerization. However, its accuracy rapidly
deteriorates due to a severe sign problem at larger values of $\alpha$,
especially at low temperatures. It recovers slightly at very large
values of $\alpha$. The inchworm method, while not as accurate at
small $\alpha$ near the part of parameter space corresponding to
a normal Hubbard model, generally suffers less from both an increase
in both $\alpha$ and the lowering of temperature. It becomes very
accurate at large $\alpha$, and generally converges more easily when
the system is in an insulating state.

\begin{figure}
\includegraphics{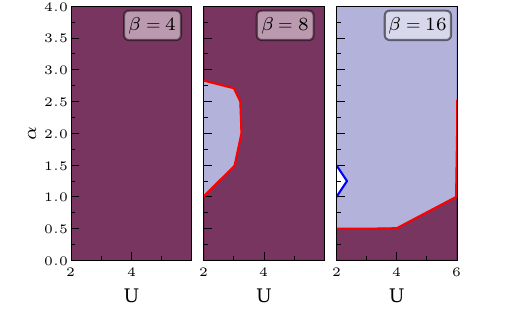}

\caption{Parameter regions where methods converge easily, as a function of
$U$ and $\alpha$. The red and blue shaded regions show where the
CT-HYB and inchworm methods, respectively, achieved errors $\Delta G$
below a threshold value of $10^{-2}$ for a given amount of computational
resources. Panels show different temperatures, decreasing from left
to right.\label{fig:methods_regions_of_applicability}}
\end{figure}

Next, in Fig.~(\ref{fig:methods_regions_of_applicability}), we map
out the dependence of the sign problem on $\alpha$, $U$ and $\beta$
at half-filling. We carry out calculations of $\Delta G$ as before.
The red and blue shaded regions denote parameters where CT-HYB and
the inchworm method, respectively, provide a value of $\Delta G$
lower than $10^{-2}$ for the given runtime. Note that this criterion
is arbitrary and the results are implementation-dependent; nevertheless,
this procedure provides a useful and intuitive description of the
limitations of the methods. In the high temperature case $\beta=4$
(left panel), both methods converge for the entire parameter regime.
At $\beta=8$ (middle panel), a region forms at $1<\alpha<3$ and
$U<4$ where the performance of CT-HYB degrades, but the inchworm
method continues to work well. As we go even lower in temperature
to $\beta=16$ (right panel), CT-HYB fails for $\alpha\gtrsim0.5$
except at large $U$. Here a parameter region where inchworm begins
to fail as well (white region) is also visible. This appears at small
$U$ and around $\alpha=1$; note that this behavior is also seen
in Fig.~\ref{fig:sign_problem_error}, where the highest peak in
$\Delta G$ for the inchworm method appears near $\alpha=1$. While
the errors in CT-HYB at these parameters make most of the parameter
space completely intractable at $\beta=16$, inchworm data could still
be obtained here given a reasonable investment in computational resources.
We expect that if it becomes necessary to perform simulations at much
lower temperatures, further algorithmic improvements will be required
even for the inchworm method.

\subsection{Studying phase transitions\label{subsec:transitions}}

We will be interested in two types of order parameter. The first distinguishes
between paramagnetic (PM) and antiferromagnetic (AFM) states, and
the second between metallic and insulating ones. These are the same
distinctions made in previous work \citep{hafermann_metal-insulator_2009},
where the entire phase diagram at half filling was obtained for $0\le U\le5$
at $\beta=10$ using CT-INT as the impurity solver. As we noted, CT-INT
has no sign problem at half filling, and converges rapidly at weak
interaction strengths. It is therefore an ideal choice for this problem,
and can serve as an excellent benchmark for the hybridization-expansion-based
inchworm method.

We will first discuss the PM/AFM transition. One interesting finding
in Ref.~\citep{hafermann_metal-insulator_2009} was that at the limit
of low temperature and large interaction strength $U$, the system
undergoes a transition from AFM to PM as $\alpha$ increases past
$\sqrt{2}$. It was suggested that this is due to the transition happening
when the effective intra-layer Heisenberg exchange coupling, $J^{\prime}\sim\frac{t^{\prime2}}{U}$,
is twice the size of its inter-layer counterpart $J\sim\frac{t^{2}}{U}$.
This, because every dimer has one $J^{\prime}$ coupling and two $J$
couplings. This low energy argument is correct at low temperatures,
but can be expected to break down at higher ones. In particular, for
larger values of $U$ we might expect this breakdown to occur when
the temperature is comparable to the larger of $J^{\prime}$ and $J$.

\begin{figure}[t]
\includegraphics{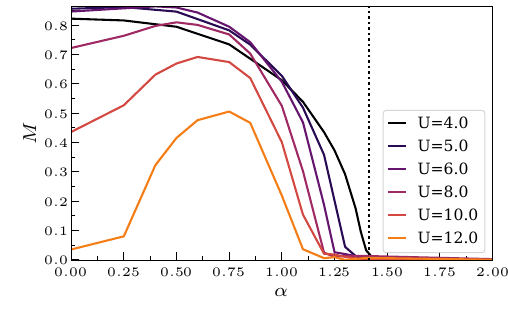}.

\caption{Zero-field magnetization at $\beta=10$ and various values of $U$,
plotted as a function of $\alpha$. The theoretical low temperature
limit for the AFM--PM transition \citep{hafermann_metal-insulator_2009}
is shown as a vertical dashed line at $\alpha=\sqrt{2}$. The line
widths indicate a confidence interval given by the absolute value
of the difference between the last two DMFT iterations.\label{fig:hafferman_break}}
\end{figure}

In Fig.~\ref{fig:hafferman_break} we plot the zero-field magnetization
predicted from our inchworm-based DMFT calculations at $\beta=10$
and half filling, for a range of $U$ values. This is defined simply
as
\begin{equation}
M=\left\langle n_{\uparrow}^{a}+n_{\downarrow}^{b}-n_{\downarrow}^{a}-n_{\uparrow}^{b}\right\rangle ,\label{eq:magnetization}
\end{equation}
and---being a static property---can be directly obtained from inchworm
Monte Carlo calculations \citep{cohen_taming_2015}. In practice,
we apply a small staggered magnetic field during the first DMFT iteration
in order to break symmetry; then we turn it off, and allow the DMFT
self-consistency to evolve to either a symmetry-broken ($M\neq0$)
or symmetric $\left(M=0\right)$ state. In practice, our numerical
criterion for this was that the zero field magnetization, Eq.~(\ref{eq:magnetization}),
be over an arbitrary threshold value, $M_{\mathrm{threshold}}\equiv0.03$.

The $U=4$ result is consistent with results from Ref.~\citep{hafermann_metal-insulator_2009},
where similar data is shown. The phase transition indeed occurs at
$\alpha\approx\sqrt{2}$, as expected from the low energy theory.
We also present a sequence of curves at higher values of $U$, still
at the same temperature. It is clear that the value of $\alpha$ at
which the transition occurs at this temperature decreases with $U$.
At $\alpha=0$ the system is a normal Hubbard model, and the decrease
in magnetization at larger values of $U$ is related to the known
decrease in the Curie temperature, which is expected to shift inversely
with $U$ \citep{georges_physical_1993,van_dongen_thermodynamics_1991,jarrell_hubbard_1992}.
At large values of $U\gtrsim6$, magnetization increases with $\alpha$
before decreasing again, showing that an intermediate inter-layer
coupling can either suppress or enhance magnetic properties.

The second order parameter we will discuss is an approximation for
the spectral function at the chemical potential, and therefore a proxy
for metallicity:
\begin{equation}
A\left(\omega=\mu\right)\simeq-\frac{\beta}{\pi}G\left(\tau=\beta/2\right).\label{eq:dos_at_fermi}
\end{equation}
The approximation becomes increasingly accurate at the low temperature
limit $\beta\rightarrow\infty$, but is commonly used in the DMFT
literature at finite temperature \citep{georges_dynamical_1996}.
One possible alternative is to perform analytical continuation and
define the transition according to the formation of a gap, but this
requires additional assumptions. A more rigorous, but challenging,
route is to solve the problem on the real axis using nonequilibrium
dynamical mean field theory \citep{schmidt_nonequilibrium_2002,freericks_nonequilibrium_2006,arrigoni_nonequilibrium_2013,aoki_nonequilibrium_2014,erpenbeck_quantum_2023};
we will explore this in future work. Here, we chose to define a simple,
but arbitrary, transition threshold. When $-\frac{\beta}{\pi}G\left(\tau=\beta/2\right)<0.25$,
we refer to the system as insulating; otherwise, we refer to is as
metallic. This definition should be seen as qualitative.

\begin{figure}
\includegraphics{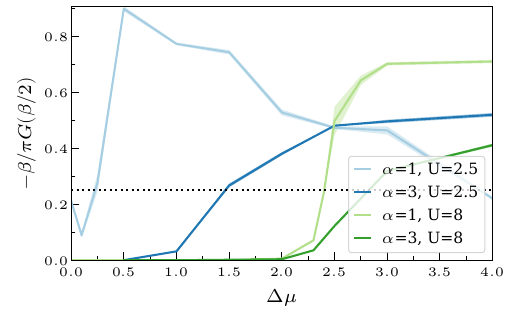}

\caption{Proxy for metallic behavior as a function of the chemical potential
shift $\Delta\mu$, at inverse temperature $\beta=16$. The dotted
horizontal black line denotes the (arbitrarily chosen) threshold value
we use to indicate the transition from an insulating to a metallic
state. The line widths indicate a confidence interval given by the
absolute value of the difference between the last two DMFT iterations.\label{fig:metallic-order-parameter}}
\end{figure}

Fig.~\ref{fig:metallic-order-parameter} shows how this definition
can be used to investigate the metal--insulator transition that occurs
when the system is doped. The metallic order parameter from Eq.~(\ref{eq:dos_at_fermi})
is shown as a function of the chemical potential, at two values of
the interaction strength $U$ and two values of the hopping ratio
$\alpha$. In all cases, higher doping takes the system from the insulating
to the metallic regime. Increasing both $U$ and $\alpha$ drives
the transition to higher chemical potentials. When $U=2.5$ and $\alpha=1$,
we can observe a drop in the order parameter at larger chemical potentials
and an eventual reentrance into insulating behavior due to the reduction
of free charge carriers. The latter will happen at high enough chemical
potential shift for all parameters.

\subsection{Phase diagram\label{subsec:phase-diagram}}

\begin{figure}
\includegraphics{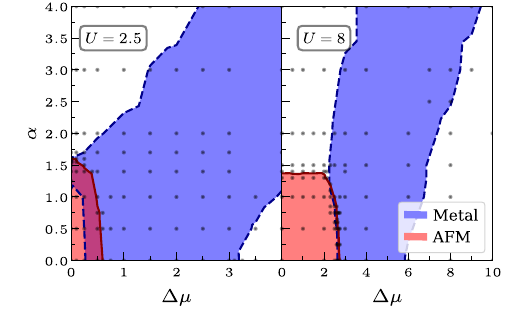}\caption{Phase diagram of the bilayer Hubbard model at $\beta=16$ as function
of the ratio between the hopping terms $\alpha=t'/t$ and the doping
$\Delta\mu$. The left and the right panel has coulomb onsite interaction
$U$ of $2.5$ and $8$, respectively. Dots indicate parameter combinations
where a calculation was made, and bilinear interpolation is used to
construct the curves as described in \subsecref{transitions}.\label{fig:phase_diagram} }
\end{figure}

Using the order parameters and thresholds discussed in \subsecref{transitions},
we performed a set of calculations at $U=2.5$ and $U=8$, mapping
out the phase diagram of the model at each of these interaction strengths
and temperature $\beta=16$ with respect to the ratio $\alpha$ between
inter-layer and intra-layer hoppings; and the chemical potential shift
from half filling, $\Delta\mu=\mu+\frac{U}{2}$. Due to symmetry considerations,
it is sufficient to examine the $\Delta\mu>0$ part of the diagram.
The two maps are presented in the left and right panels of Fig.~\ref{fig:phase_diagram}.
The shaded red region in each denotes parameters where we found the
system to be antiferromagnetic. Similarly, the blue region denotes
parameters where we predict metallic behavior. Four combinations of
order parameters are possible, embodying four distinct phases: the
AFM insulator, AFM metal, PM insulator and PM metal.

We begin by considering previously available results. The $\alpha=0$
case (bottom edge of both panels in Fig.~\ref{fig:phase_diagram})
corresponds to the normal Hubbard model. The behavior there is consistent
with that known in the literature \citep{kajueter_doped_1996}: for
$U=2.5$, the system is in the AFM state for $\Delta\mu\lesssim0.5$.
The metallic regime begins when $\Delta\mu\approx0.2$ and ends at
$\Delta\mu\approx0.5$. At $U=8$, the AFM regime expands while the
insulating regime shifts to larger values of $\Delta\mu$.

Next, we consider the $\Delta\mu=0$ case (left edge of both panels
in Fig.~\ref{fig:phase_diagram}), which was explored by Hafermann
et al. in Ref.~\citep{hafermann_metal-insulator_2009}. At $U=2.5$,
the system is an AFM Mott insulator at $\alpha=0$. Increasing $\alpha$
drives it first to an AFM metallic phase, then to a PM band insulator
state. Depending on the thresholds and temperature, a small PM metallic
region may exist at $\alpha\approx1.7$ \citep{hafermann_metal-insulator_2009}.
At $U=8$, however, the system goes directly from an AFM insulator
to a band insulator at $\alpha\approx\sqrt{2}$; this behavior was
also found in Ref.~\citep{hafermann_metal-insulator_2009} at $U\apprge4$.
Note also the difference from the behavior in Fig.~\ref{fig:hafferman_break},
where $\beta=10$.

The rest of the phase diagram (everywhere else in the two panels of
Fig.~\ref{fig:phase_diagram}, where neither $\alpha=0$ nor $\Delta\mu=0$)
is where both DMFT calculations based on the standard CT-HYB, CT-INT
or CT-AUX methods face sign problems unless special symmetries can
be taken advantage of, as in Refs.~\citep{najera_resolving_2017,najera_multiple_2018}.
Our CT-HYB-based inchworm method works well in the entire parameter
space, without any need for symmetrization, and can be used to obtain
both the metallic and AFM order parameters. For both values of $U$,
an AFM regime appears at small $\alpha$ and $\Delta\mu$. Increasing
$U$ extends this regime in $\Delta\mu$, while simultaneously contracting
it in $\alpha$. The metallic behavior appears at intermediate values
of $\Delta\mu$ and is shifted approximately linearly in $\alpha$.
The metallic region always appears to be simply connected. In particular,
at small $U$ the AFM metallic state on the $\Delta\mu=0$ line is
connected to its counterpart on the $\alpha=0$ line. As a result
of this, at small $U$ the metallic and AFM regimes overlap in a narrow
band surrounding the AFM insulator at the origin, but this overlap
shrinks with increasing $U$ as the metallic phase recedes from the
$\Delta\mu=0$ line.

We emphasize that the particular size and shape of the regions and
overlaps depends to some degree on our choice of order parameters
and thresholds, but we expect the general physical trends to be robust.

\begin{figure}

\includegraphics{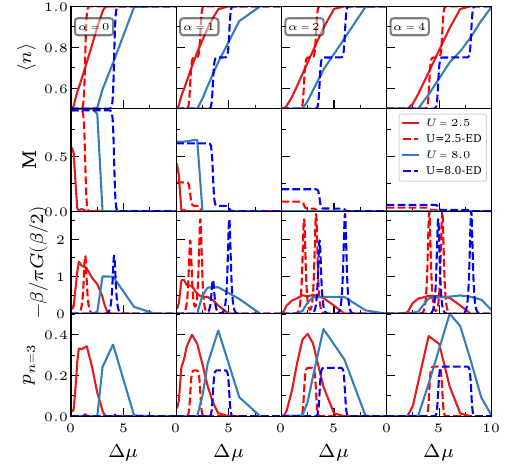}\caption{The average occupation (top row), staggered magnetization (second
row), metallic order parameter (third row), and probability of occupying
states with $3$ electrons in the unit cell (bottom row); all as a
function of the chemical potential shift $\Delta\mu$, for two values
of the coulomb interaction (red and blue curves) and for four values
of the ratio $\alpha$ between inter-band and intra-band hopping amplitudes
(in different columns). Solid lines denote the full DMFT solution
on the Bethe lattice with inchworm Monte Carlo as the impurity solver.
Dashed lines are ED calculations for the two-dimer model (see text).\label{fig:local_order_parameters}}
\end{figure}

Much of the dependence of physical behavior on doping can stem from
local mechanisms like energetics and population switching effects
within a unit cell. It can therefore be understood to some degree
by considering a substantially simplified model: two adjacent dimers
(here a dimer refers to two coupled orbitals, one of which is in each
layer). This model contains only four spin-half orbitals, and can
easily be solve by exact diagonalization (ED); it also becomes equivalent
to the lattice model at the fully dimerized limit $\alpha\rightarrow\infty$,
and bears some resemblance to a DMFT calculation with a minimal discrete
bath, at least with paramagnetic boundary conditions. Since no spontaneous
symmetry breaking can occur in the ED solution without a self-consistency
condition, we solve the system in the presence of a small staggered
field $h=0.1$ that induces an AFM state. In Fig.~\ref{fig:local_order_parameters}
we present a detailed view of how several local observables depend
on the chemical potential, for $U=2.5$ and $U=8$ (red and blue curves,
respectively); and for the numerically exact solution on Bethe lattice
and the two-dimer model (solid and dashed lines, respectively).

The top row of Fig.~\ref{fig:local_order_parameters} shows the mean
occupancy of the orbitals within a unit cell. At $\Delta\mu=0$ the
system is half filled: there are, on average, two electrons per unit
cell. Naturally, the occupancy rises as the chemical potential increases,
and at $\Delta\mu=10$ the system becomes completely filled (4 electrons
per unit cell) at all the parameter sets shown here. In the lattice
model, the increase in occupation is continuous and mostly linear.
In the two-dimer model, it increases in a series of steps with a width
set by the temperature, but over a similar range of chemical potentials.

The second row of Fig.~\ref{fig:local_order_parameters} shows the
AFM order parameter $M$. It is clear that as the occupation increases,
magnetic behavior is suppressed. This is not surprising when considering
that the half-occupied local state $\left|\uparrow\right\rangle _{a}\left|\downarrow\right\rangle _{b}$
has the maximal AFM response. In the two-dimer model, the drop in
magnetism is associated with the first jump in population. However,
it is clear that the two-dimer model does not fully capture the physics
of the AFM--PM transition, even with the addition of a symmetry-breaking
field.

We now consider the metal--insulator transition, which is captured
by the order parameter of Eq.~(\ref{eq:dos_at_fermi}). This is plotted
in the third row of Fig.~\ref{fig:local_order_parameters}. While
the two-dimer always produces sharp peaks in the proxy for the density
of states, these peaks do appear in the approximate region where the
broadened density of states appears in the lattice model. This region
appears when the system begins to exit the half-occupied state, which
characterizes a Mott insulator; and disappears when it enters the
fully-occupied state, which characterizes a band insulator.

Finally, the bottom row of Fig.~\ref{fig:local_order_parameters}
shows the probability of occupying metal-like local states with 3
electrons. In the lattice model, this probability appears and disappears
in tandem with the presence of a metallic order parameter. In the
two-dimer model, the association between these observables is still
apparent, especially at larger values of $\alpha$; but both the magnitude
and extent in chemical potential where this state is likely are significantly
underestimated. The large discrepancy between the two-dimer and lattice
models in all panels on the bottom row of Fig.~\ref{fig:local_order_parameters}
illustrates the fact that metallic behavior, in particular where it
is associated with Kondo correlations, is where the analogy between
these two models breaks down most noticeably.

\section{Conclusions\label{sec:Conclusions}}

We presented a study of the infinite-dimensional bilayer Hubbard model
on the Bethe lattice. The dynamical mean field approximation is exact
for this model, given a procedure for solving the associated multiorbital
quantum impurity model. We showed that the inchworm quantum Monte
Carlo method enables this solution in parameter regimes where other
methods fail due to severe sign problems. This allowed us to obtain
the model's phase diagram outside of half-filling.

Our results show that at combinations of temperature and interaction
strength where the half-filled model has a metallic regime, it merges
with the corresponding metallic regime in the normal Hubbard model,
where the two layers are isolated from each other. It can then overlap
with the antiferromagnetic regime that exists near half-filling and
at weak inter-layer coupling. At stronger interaction strengths the
metallic regime is pushed towards higher chemical potentials. We also
showed that most, though not all, of the qualitative properties of
the phase diagram can be understood from the local energetics and
state population probabilities within a unit cell.

In addition to shedding light on the properties of the doped bilayer
Hubbard model, our work illustrates some of the limitations of standard
quantum Monte Carlo methods in the study of multiorbital strongly
correlated electron physics. It then highlights inchworm techniques
as a way around such limitations, showing that they enable access
to parameter regimes that were previously difficult to simulate. We
expect the methodology introduced here may be useful for many other
situations where sign problems have limited our ability to answer
important scientific questions regarding complex quantum materials.
\begin{acknowledgments}
This research was supported by the ISRAEL SCIENCE FOUNDATION (Grants
No. 2902/21 and No. 218/19) and by the PAZY foundation (Grant No.
318/78).
\end{acknowledgments}

\bibliographystyle{apsrev4-1}
\addcontentsline{toc}{section}{\refname}\bibliography{library}

\end{document}